%% file: Master.tex
\documentclass[10pt,twoside,twocolumn]{book}
\usepackage{jrjc}

\begin{document}

%%%%%%%%%%%%%%%%%%%%%%%%%%%%%%%%%%%%%%%%%%%%%%%%%%%%%%%%%%%%

%\rjcinclude{session}{DinhThi}

%
  {%                            in local group to allow private
   %                            redefinitions
    \cleardoublepage%           make sure we start on the right...
    \subimport{session/DinhThi/}{DinhThi}%
  }
  % if you do not want to include a photo of you

%%%%%%%%%%%%%%%%%%%%%%%%%%%%%%%%%%%%%%%%%%%%%%%%%%%%%%%%%%%%

\end{document}

%% file: session/DinhThi/DinhThi.tex
\title{ Confronting the nucleonic hypothesis with current neutron star observations from GW170817 and  PSR J0740+6620}
\author{H. Dinh Thi, C. Mondal,
F. Gulminelli}
\institute{Laboratoire de Physique Corpusculaire (LPC), CNRS, ENSICAEN, UMR6534, Universite de Caen Normandie, F-14050 Caen Cedex, France }

\abstract{The nuclear matter equation of state is relatively well constrained at sub-saturation densities thanks to the knowledge from nuclear physics. However, studying its behavior at supra-saturation densities is a challenging task. Fortunately, the  extraordinary progress recently made in observations of neutron stars and neutron star mergers has provided us with unique opportunities to unfold the properties of  dense matter. Under the assumption that nucleons are the only constituents of neutron star cores, we perform a Bayesian inference using the so-called meta-modeling technique with a nuclear-physics-informed prior. The latest information from the GW170817 event by the LIGO-Virgo Collaboration (LVC) and from the radius measurement of the heaviest known neutron star PSR J0740+6620 by the Neutron Star Interior Composition Explorer (NICER) telescope  and X-ray Multi-Mirror (XMM-Newton) are taken into account as likelihoods in the analysis. The impacts of different constraints  on the equation of state as well as on the predictions of neutron star properties  are discussed. The obtained posterior reveals that all the current observations are fully compatible with the nucleonic hypothesis. Strong disagreements between our results with future data can be identified as a signal for the existence of exotic degrees of freedom.}
\maketitle
\section*{Introduction}
The  knowledge of equation of state (EoS)  at several times the nuclear saturation density ($\rho_{\rm sat} = 2.8 \times 10^{14}$ g cm$^{-3}$) is not yet accessible in terrestrial laboratories. Nonetheless, these densities can be explored in neutron star (NS) cores \cite{hpy2007}. Together with the growth of multi-messenger astronomy, over the last few years, we have witnessed several breakthroughs in NS observations. For instance, astrophysicists are able to determine heavy pulsar masses  with high precision via radio timing, e.g. PSR J0348+0432 ($M = 2.01\pm 0.04 M_{\odot}$) \cite{Antoniadis2013} and  PSR J0740+6620 ($M = 2.08\pm 0.07 M_{\odot}$) \cite{Fonseca2021}, where $M_{\odot}$ is solar mass. Moreover, in August 2017, gravitational waves from the merger of a  binary NS system (GW170817)  were detected by LVC, delivering the very first information about the tidal deformability of NS \cite{ligo1, ligo2, ligo3}. Additionally, the development of X-ray timing telescope brings us information about the joint mass-radius distributions of the millisecond pulsars PSR J0030+0451 (NICER data) \cite{nicer1, nicer2} and PSR J0740+6620 (NICER+XMM-Newton data) \cite{nicer3, nicer4}. Due to the one-to-one correspondence between the EoS and NS static observables \cite{Hartle1967}, these measurements together with the  upcoming data \cite{ligo4} can be transformed into valuable information on matter at extreme conditions that cannot be reproduced on Earth.  Thus, they are expected to be very promising tools to uncover the open questions about dense matter, such as whether or not exotic degrees of freedom exist in the cores of NS \cite{Oertel2017}.

In this work, we assume that the only baryonic constituents of  NS cores are  nucleons, which are in weak equilibrium with electrons and muons. Under this hypothesis, the energy functional is described using the so-called meta-modeling technique \cite{metamodel_1, metamodel_2}.  With the metamodel, the nuclear matter energy can be characterized by empirical parameters through a simple parameterization. In our Bayesian inference, the range of these parameters in the prior distribution is chosen such that they are compatible with the current nuclear physics knowledge \cite{metamodel_1, metamodel_2, CarreauEPJA, Hoa2021, HoaUniverse21}, hence the name nuclear-physics-informed prior. This analysis can be considered as a way of transforming information from nuclear physics experiments and calculations as well as from astrophysical observations into empirical parameters in order to guide the elaboration of phenomenological and microscopic nuclear models. Additionally, it can be used as a null hypothesis to search for exotic degrees of freedom.

In the next section, we briefly recall the formalism described in Ref. \cite{HoaUniverse21}. Then, we discuss the posterior results from the Bayesian analysis. Finally, we present our conclusions. The main results of this study are published in Ref. \cite{HoaUniverse21}.

\section*{Method}
\subsection*{Meta-modeling technique}
The meta-modeling for the EoS, proposed in Margueron et al. (2018) \cite{metamodel_1}, is inspired from the Taylor expansion around the saturation density $n_{\rm sat}$ of nuclear matter energy. It was shown in Ref. \cite{metamodel_1} that any nucleonic EoS can be reproduced satisfactorily by truncating the series expansion at the fourth order. The energy per nucleon at density $n = n_n + n_p$ and asymmetry $\delta =(n_n-n_p)/n$, therefore, can be written as
\begin{equation}
e(n,\delta) \approx \sum_{m=0}^4 \frac{1}{m!} \left ( \left. \frac{d^m e_{\rm sat}}{d x^m} \right|_{x=0} 
+ \left.  \frac{d^m e_{\rm sym}}{d x^m} \right|_{x=0}\delta^2\right ) x^m ,
\label{eq:energy_per_nuc}
\end{equation} 
 where $n_n$, $n_p$ are the neutron and proton densities, $x = (n-n_{\rm sat})/(3n_{\rm sat})$, $e_{\rm sat}$ is the energy per nucleon of symmetric matter, and $e_{\rm sym}$ is called the symmetry energy, which is the difference between the energy per baryon of pure neutron matter and that of symmetric matter. In addition, the energy functional also includes the isoscalar effective mass and effective mass splitting in the kinetic energy and a parameter governing the energy functional at the zero-density limit. Thus, in total, the bulk energy is characterized by 13 parameters. To describe NS crusts, we complement these bulk parameters with 5 surface and curvature parameters  within a  compressible liquid drop model \cite{Hoa2021}.  For each set of uniform matter parameters, these surface and curvature parameters are obtained by the optimal fit to the experimental Atomic Mass Evaluation 2016 (AME2016) nuclear mass table \cite{ame2016}. 

\section*{Statistical analysis}
Here, we recall the main points in the Bayesian analysis performed in Ref. \cite{HoaUniverse21}.

According to the  Bayes' theorem, the posterior probability of the bulk parameter set $\bf X$ at given constraints $\mathbf c $ can be written as:
\begin{equation}
	P({\mathbf X}|{\mathbf c})=\mathcal N   P({\mathbf X}) \prod_k P(c_k|{\mathbf X}),
\end{equation}
with  $\mathcal N$ being a normalization factor. $P({\mathbf X})$  and $P(c_k|{\mathbf X})$ are called prior and likelihood, respectively. The constraints implemented in this work take into account both information at low density from nuclear mass measurements in the AME2016 mass table \cite{ame2016} and many-body perturbation theory (MBPT) calculations \cite{Drischler2016} and at high density from astrophysical observations \cite{Antoniadis2013, ligo3, nicer2, nicer4}. 

Considering an observable $Y$, the posterior distribution of $Y$ can be calculated as
\begin{equation}
	P(Y|{\mathbf c})=\prod_{k=1}^N\int_{X_k^{min}}^{X_k^{max}}dX_k \, P({\mathbf X}|{\mathbf c}) \delta\left (Y-Y({\mathbf X})\right ).
	\label{eq:Ypost}
\end{equation}
In Equation \ref{eq:Ypost}, $N=13$ is the number of bulk parameters, $X_k^{\rm min(max)}$ is the parameter minimum (maximum) value   chosen  based on the current nuclear physics knowledge \cite{metamodel_1, CarreauEPJA}, and  $Y({\mathbf X})$ is the value obtained with the parameter set ${\mathbf X}$ of the observable $Y$.

With the aim of studying the effects of different constraints on the nuclear-physics-informed prior, we present our results in four distributions as follows:

\begin{enumerate}
\item {\it Prior}: Models in the prior distribution are compatible with nuclear physics and represent the most general predictions within the nucleonic hypothesis. The weight of each parameter set ${\mathbf X}$ is determined by the quality of the optimization of the surface and curvature parameters to fit the nuclear masses in the AME2016 table \cite{ame2016}. 
\item {\it LD}:  The information from ab-initio calculation is included. Particularly,  models are selected  by a pass-band filter which is the chiral EFT calculation for the energy per nucleon of symmetric  matter and pure neutron matter in Ref. \cite{Drischler2016}. The filter is applied in the density interval $\left[ 0.02, 0.2\right]$ fm$^{-3}$.
\item {\it HD+LVC}: In this distribution, models are required to satisfy the following criteria: causality, thermodynamic stability, and non-negative symmetry energy at all densities. Then, the weight of each parameter set ${\mathbf X}$ is evaluated using two measurements: (1) mass measurement of  PSR J0348+0432 \cite{Antoniadis2013}, which is interpreted as a cumulative Gaussian distribution function  with mean value $\mu = 2.01$ and standard deviation $\sigma = 0.04$, in the unit of $M_{\odot}$. This condition is used on the maximum NS mass obtained for each EoS; (2) joint distribution of tidal deformability and mass ratio inferred from the GW170817 by LVC as in Refs. \cite{ligo3, dataligo}.
\item {\it All}: This includes  the constraints mentioned above together with the likelihoods from the two mass-radius measurements from Refs. \cite{nicer2,nicer4}. This separation allows us to identify the effect from the new radius measurement of PSR J0740+6620 \cite{nicer4}.
\end{enumerate}
The model rejection (or acceptance)  rate depends on the filter implemented. In each figure shown in the following section, similar statistics are used in the four distributions to make sure that the difference among them originate from the physical constraints. Moreover, the chosen statistics is checked to be sufficient so that convergence can be reached.

\section*{Results and Discussions}
\subsection*{NS crustal and global properties}
As we have shown in  Ref. \cite{HoaUniverse21}, the ab-initio  nuclear physics  calculation and the astrophysical data have distinct impacts on NS properties. To be more specific, the tightness of these constraints depends upon whether  the crustal or global properties are considered.

\begin{figure}[htbp]
\begin{center}
\includegraphics[width=5.5cm,keepaspectratio]{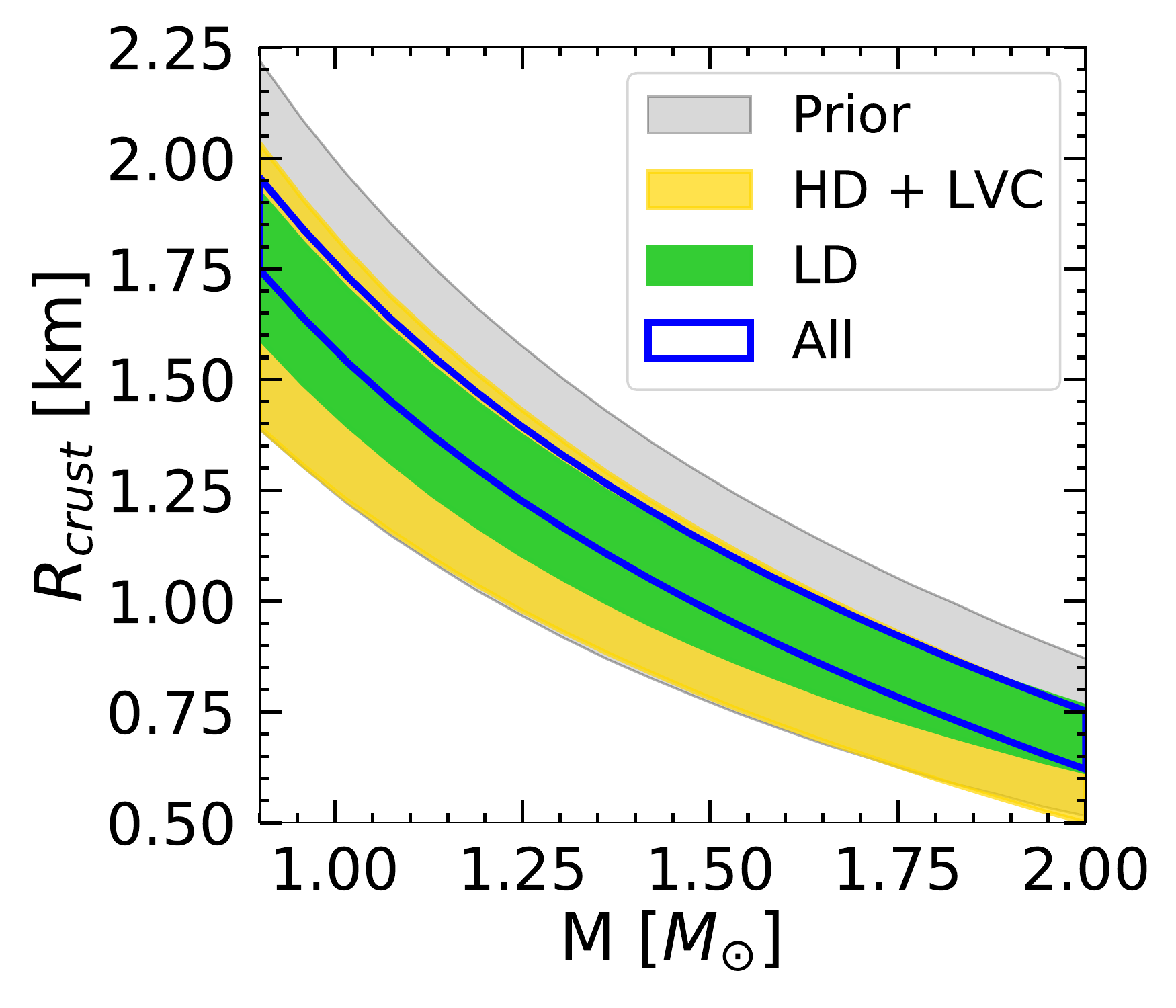}
\end{center}
\caption{68$\%$ confidence intervals of crustal thickness $R_{\rm crust}$ as a function of NS mass $M$ in the four distributions: Prior (gray), HD+LVC (yellow), LD (green), and All (blue). }
\label{Rcrust}
\end{figure}

\begin{figure}[htbp]
\begin{center}
\includegraphics[width=7.7cm,keepaspectratio]{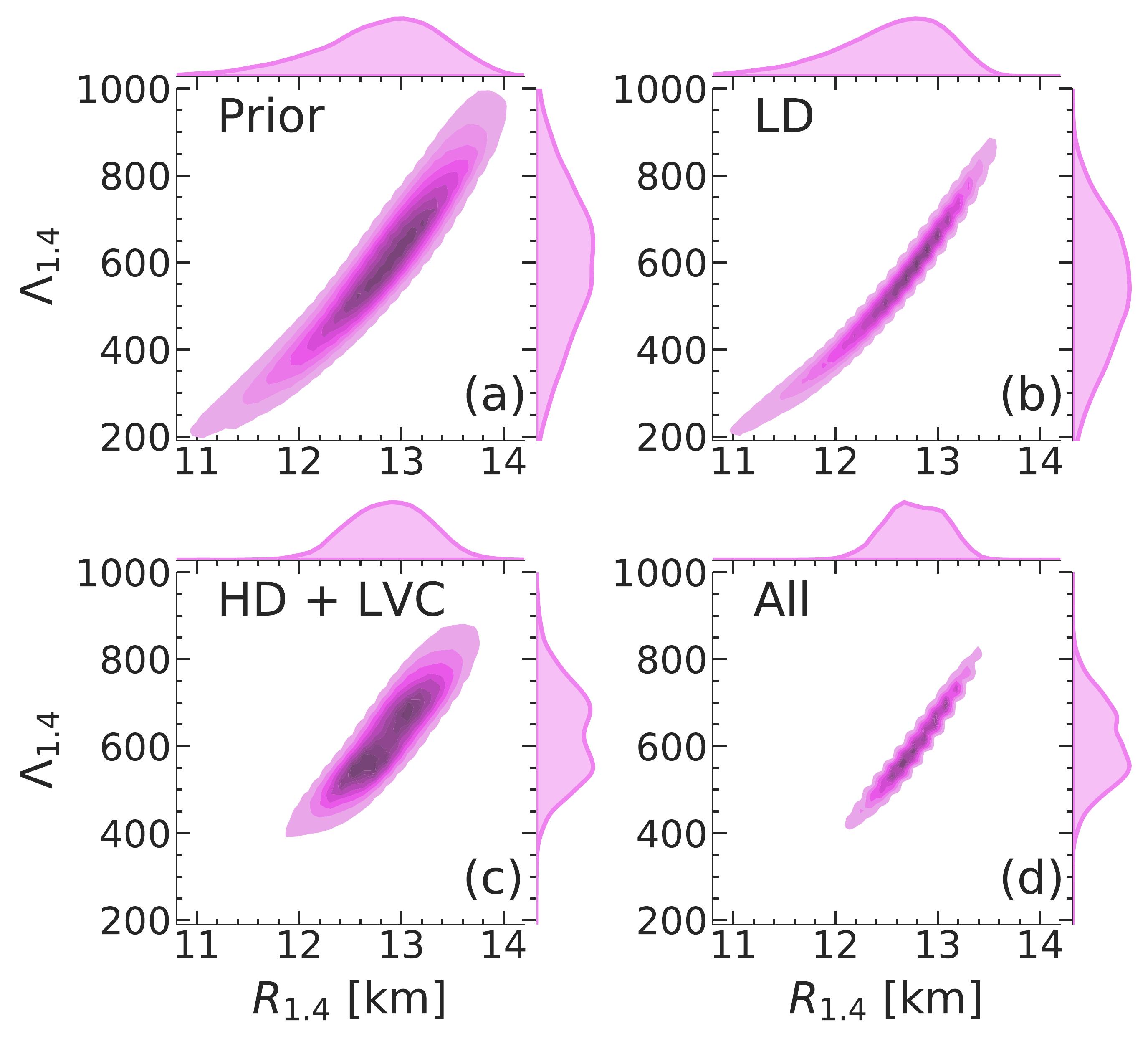}
\end{center}
\caption{Joint probability density plots of NS dimensionless tidal deformability $\Lambda$ and NS radius $R$ at $M = 1.4 M_{\odot}$ in the four distributions: Prior (panel \textbf{a}), LD (panel \textbf{b}), HD+LVC (panel \textbf{c}), and All (panel \textbf{d}).  }
\label{Lambda14_R14}
\end{figure}

 For the crustal properties, Figure \ref{Rcrust} displays the $68\%$ confidence intervals (CI) of the NS crust thickness $R_{\rm crust}$ as a function of NS mass $M$ in the prior and three posterior distributions. By construction, this quantity has a direct correlation to the location of the crust-core transition. In Ref. \cite{HoaUniverse21}, it was shown that the chiral EFT calculation strictly constrains both the crust-core transition density, $n_{CC}$, and pressure, $P_{CC}$, while the astrophysical observations only disfavor very high $P_{CC}$. Therefore, we  expect that the results of $R_{\rm crust}$ should reflect very similar effects. Indeed, as it can be seen in Figure 1,  the $68\%$ CI band in the ``LD'' distribution (green) is evidently thinner than the  one in the ``Prior'' distribution (gray). On the other hand, in the ``HD+LVC'' distribution (yellow), only the upper limit of $R_{\rm crust}$ is impacted. As it was already pointed out in Ref. \cite{HoaUniverse21}, these models which result in high $R_{\rm crust}$, or equivalently high $P_{CC}$, are associated to the violation of at least one of the following requirements: causality, thermodynamics stability, and non-negative symmetry energy. Accordingly, we can safely deduce that the constraints from GW170817  and the mass measurement of the heavy pulsar PSR J0348+0432  have negligible impacts on the crustal observables. When all the filters are combined together with the two NICER measurements (blue), we obtain a narrow band for the crust thickness, in which the effect from the chiral EFT calculation is dominant. The final relative uncertainties  in the ``All'' distribution is  up to  $\sim 10\%$. The study of NS crustal properties is important because  of the crust role in explaining pulsar phenomena, such as the  ``glitches'' \cite{glitch}. Thus, further constraints in the low-density region is of great relevance.

For the global properties, we plot in Figure \ref{Lambda14_R14} the joint distributions of NS dimensionless tidal deformability and radius at the canonical mass $M = 1.4 M_{\odot}$, $\Lambda_{1.4}$ and $R_{1.4}$. In each panel, on the axes we display the marginalized probability density distributions of $\Lambda_{1.4}$ and $R_{1.4}$, while the plot in the center shows the correlation between them. From panel \textbf{b}, it is clear that the chiral EFT calculation   constrains only slightly the upper limits of these two quantities. In addition, the correlation between $R_{1.4}$ and $\Lambda_{1.4}$ is enhanced in this case. In the ``HD+LVC'' distribution (panel \textbf{c}), the uncertainties in both $R_{1.4}$ and $\Lambda_{1.4}$ are significantly reduced. Specifically, the constraint from the GW170817 event prefers soft EoS, hence lowering the upper limit of $\Lambda_{1.4}$ and $R_{1.4}$. Contrarily, the NS mass measurement favors stiff EoS. Consequently, models resulting in very low $\Lambda_{1.4}$ and $R_{1.4}$ are eliminated. The combination of these two effects in the ``HD+LVC'' distribution, therefore, leads to narrow radius and tidal deformability distributions. Finally, when all constraints are put together (panel \textbf{d}), the correlation between the two quantities becomes even more well-defined, while the uncertainties remain roughly the same as in case of the ``HD+LVC'' distribution.  This indicates that even for global properties of NS the results are insensitive to the constraints from the NICER mass-radius measurements. We will discuss this point more thoroughly in the next section. Concerning the correlation, in Ref. \cite{HoaUniverse21}, we found the Pearson correlation coefficient between $R_{1.4}$ and $\Lambda_{1.4}$ to be 0.97, which is almost a perfect positive linear relationship. This linear correlation is also discussed in several works in literature \cite{Malik2018,Fattoyev, Annala2018,Lourenco2019}.
%figure

\subsection*{Comparison with NS observations from NICER and LVC}
The validity of the nucleonic hypothesis can be checked by confronting the marginalized posteriors resulted from our Bayesian analysis with recent astrophysical observations.

\begin{figure}[htbp]
\begin{center}
\includegraphics[width=8.5cm,keepaspectratio]{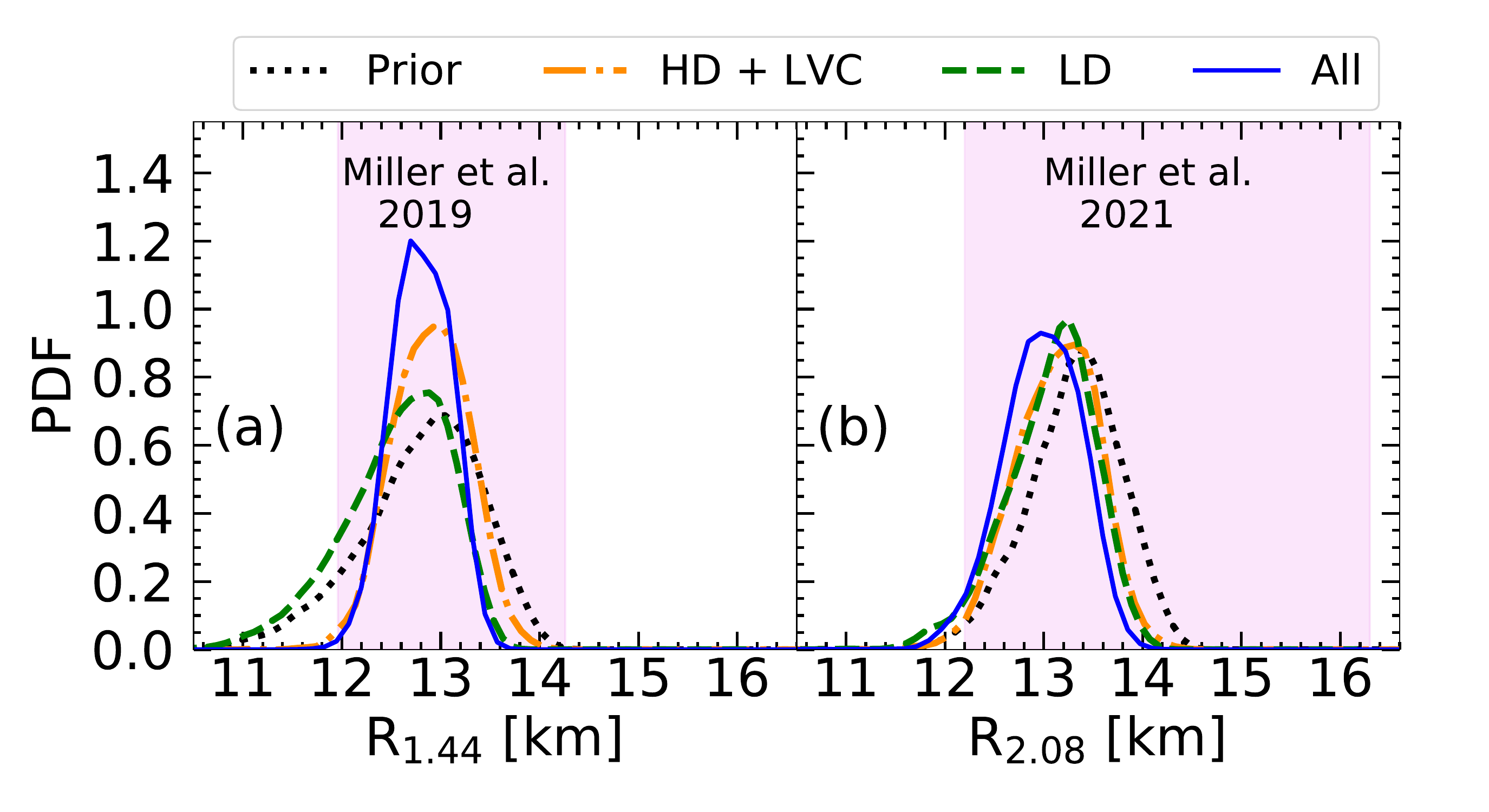}
\end{center}
\caption{Probability density distributions of NS radii in the four distributions: Prior (black dotted lines), HD+LVC (orange dash-dotted lines), LD (green dashed lines), and All (blue solid lines). Panel \textbf{a}: $M = 1.44 M_{\odot}$. The violet shaded rectangle shows radius measurement at $68\%$ confidence interval of the pulsar PSR J0030+0451  by \cite{nicer2}. Panel \textbf{b}: $M = 2.08 M_{\odot}$. The violet shaded rectangle shows the radius measurement at $68\%$ confidence interval of the pulsar PSR J0740+6620  by \cite{nicer4}.}
\label{radius}
\end{figure}

\begin{figure}[htbp]
\begin{center}
\includegraphics[width=8.9cm,keepaspectratio]{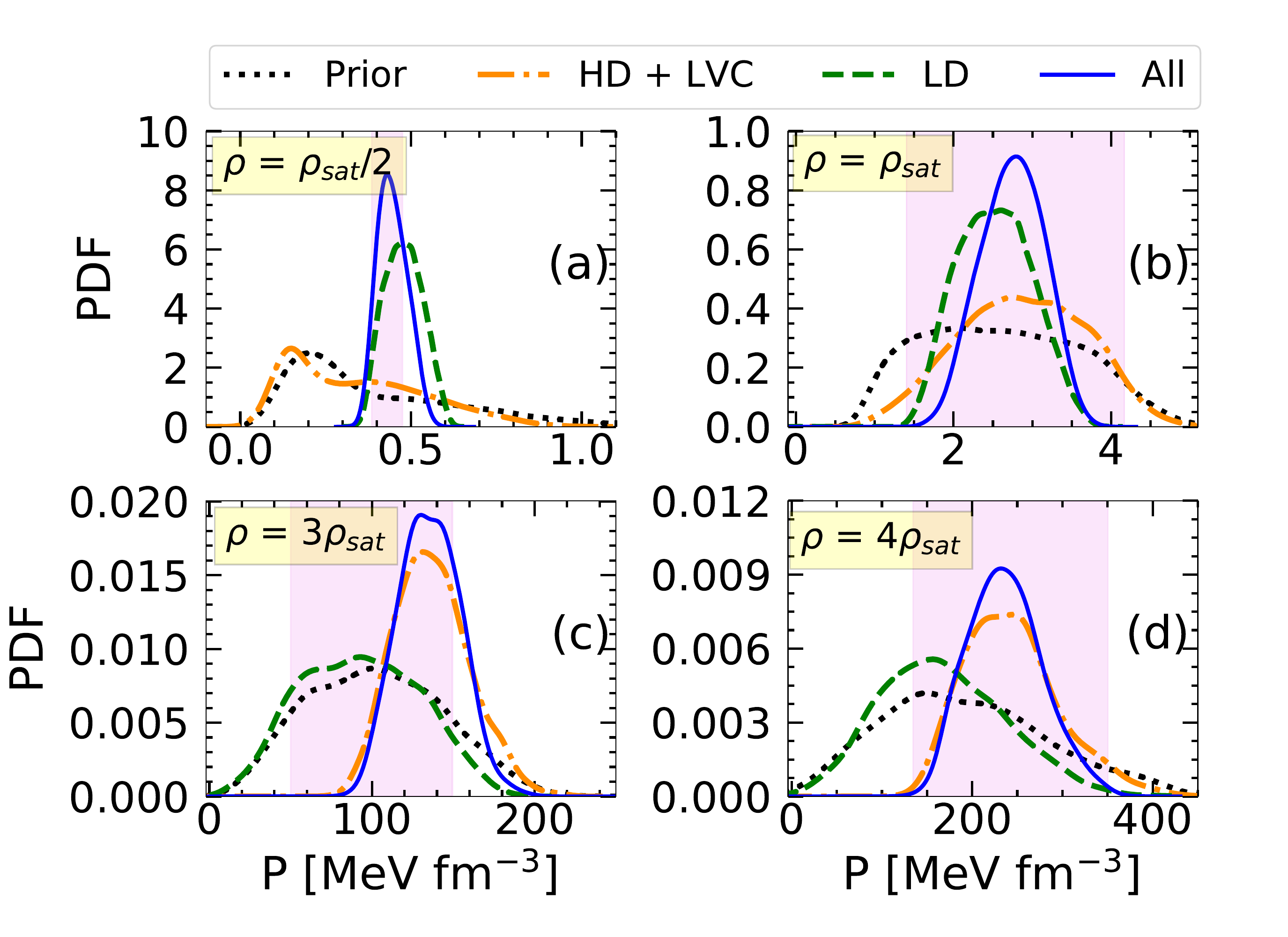}
\end{center}
\caption{Probability density distributions of  pressure in the four distributions at densities $\rho = \rho_{sat}/2$ (panel \textbf{a}),  $\rho = \rho_{sat}$ (panel \textbf{b}), $\rho = 3\rho_{sat}$ (panel \textbf{c}), and $\rho = 4\rho_{sat}$ (panel \textbf{d}). The violet shaded rectangles display the corresponding pressure at $90\%$ confidence interval inferred from the  GW170817 event  by Abbott et al. 2018 \cite{ligo2}.}
\label{pressure}
\end{figure}

Figure \ref{radius} displays the probability density distributions of NS radii at $M = 1.44 M_{\odot}$ and $M = 2.08 M_{\odot}$, which are respectively the masses (at median values) of pulsars PSR J0030+0451 \cite{nicer2} and PSR J0740+6620 \cite{nicer4}. Considering the first case (panel \textbf{a}), we can see that the ``HD+LVC'' distribution (dash-dotted orange line) is noticeably narrower than the ``LD'' distribution (dashed green line). This is  similar to what we have discussed in Figure \ref{Lambda14_R14}. As a result, the ``All'' distribution (blue solid line) is chiefly constrained by the NS mass and tidal deformability measurements from Refs. \cite{Antoniadis2013, ligo3}. On the other hand, in panel \textbf{b}, we do not observe noteworthy difference among the four distributions. This is because in this case all EoSs are required to be hard enough to support the mass 2.08$M_{\odot}$. Consequently,  the lower limits of $R_{2.08}$ in all distributions are strongly restricted as very soft EoSs are filtered out. Furthermore, this condition makes the requirement for the maximum NS mass in the ``HD+LVC'' distribution redundant. Therefore, the deviation between the ``Prior'' and the ``HD+LVC'' distributions appears mainly due to the constraint from GW170817 by LVC. Although the effects of the ab-initio nuclear physics calculation and the tidal deformability measurement  on the upper limits of $R_{2.08}$
look  similar (see the green dashed line and  dash-dotted orange line in panel \textbf{b}), these two constraints dominate very different regions in NS properties. Particularly, the former governs the crust thickness, whereas the latter determines the global radius, as we have already seen in Figures \ref{Rcrust} and \ref{Lambda14_R14}.

In both cases, our posterior distributions are in excellent agreement with the mass-radius measurements from Refs. \cite{nicer2, nicer4} (violet shaded rectangles). Here, the two referred results are presented at $68\%$ CI.
From the compatibility between our predictions and NICER measurements, we can infer that these dense matter observations can still be accounted for by the nucleonic EoS. Similar conclusions were made in Refs. \cite{Pang2021,Legred2021}, where the authors computed the Bayes factors and found that the hadronic composition is favoured over the strong phase transition to quark matter. However, notice that this does not eliminate the possibility of having exotic degrees of freedom. In order to have a conclusive establishment in this regard, we need  more stringent measurements.

In Figure \ref{pressure}, we plot the probability density distributions of pressure at four chosen densities: $\rho = \rho_{\rm sat}/2$ (panel \textbf{a}), $\rho = \rho_{\rm sat}$ (panel \textbf{b}), $\rho = 3\rho_{\rm sat}$ (panel \textbf{c}), and $\rho = 4\rho_{\rm sat}$ (panel \textbf{d}). As it is mentioned in the previous section, the prior distribution in our Bayesian analysis is generated in consistency with nuclear physics experiments and theory \cite{metamodel_1}. The prior marginalized distributions for pressure in Figure \ref{pressure} are presented by black dotted lines. It is clear from the figure that our ``Prior'' distributions encompass the $90\%$ CI inferred from GW170817 by LVC \cite{ligo2} (violet shaded rectangles). At low densities (panels \textbf{a} and \textbf{b}), the pressure is primarily constrained by the chiral EFT calculation (green dashed lines). On the contrary, at high densities, the pressure is mostly impacted by the  constraints in the ``HD+LVC'' distributions (dash-dotted orange lines). These results show that the low-density part of the EoS is mainly influenced by nuclear physics inputs, whereas the high-density region is significantly constrained by astrophysical data. By comparison with the posterior pressure obtained from LVC \cite{ligo2}, we can conclude our final distributions (blue solid lines) are consistent with their results. 

Last but not least, the similarity between the ``HD+LVC'' and ``All'' distributions also implies that the two NICER measurements does not have stringent impacts on the EoS and NS properties. This is due to the fact that the uncertainties from these measurements remains sizable. Additionally, these radius measurements almost entirely overlap our prior distributions (see Figure \ref{Rcrust}). These are the two factors for which the NICER measurements are not constraining.
\section*{Conclusion}
To summarize, we have studied the impacts of different constraints from nuclear physics as well as astrophysical data on NS properties and nuclear matter EoS. Under the assumption that NS cores only consist of nucleons, we have carried out a Bayesian analysis using the meta-modeling technique with a nuclear-physics-informed prior. We have found that information from nuclear physics calculation tightly constrains the low-density parts of the EoS, hence controlling the crustal properties. Conversely, the high-density regions of the EoS, and therefore the global properties of NS, are more constrained by astrophysical data. Finally, we have shown that our results agree very well with both data from LVC and NICER. As a result, the nucleonic hypothesis cannot be ruled out. 
 
 %\newpage